\documentclass[twocolumn]{elsart}
\usepackage{graphicx}

\def\gW{W}
\begin{document}
\begin{frontmatter}
\title{On the Possibility of Combustion of Neutrons into Strange Quark Matter}

\author{I.Tokareva$^{1}$} \ead{iya@tx.technion.ac.il} and \author {A.Nusser$^{1,2}$}\ead{adi@physics.technion.ac.il}

\address{$^{1}$Physics Department, Technion, Technion-city, Haifa 32000, Israel\\
$^2$ Astrophysics, Oxford University, Keble Road, Oxford OX1 3RK, UK}

\begin{abstract}
It is shown
that a conversion of neurons into strange quark matter in
neutron stars  is possible by means of a  combustion process with a well-defined front.
Conditions for the realization of a specific combustion mode,
whether deflagration, detonation, or fast combustion, are discussed
for several forms of the  equations of state of neutron and strange matters.
\end{abstract}

\begin{keyword}
Neutron stars \sep dense matter\sep combustion
\PACS 97.60.Jd \sep 95.30.Lz \sep 26.60.+c \sep 12.39.Ba
\end{keyword}
\end{frontmatter}

\section{Introduction}
Recently, the possibility of  emergence end existence of   strange
quark stars has been intensively discussed in the literature
\cite{Alcoc,Olinto}. A likely origin for strange stars is the
conversion of neutrons into strange quark matter in neutron stars
\cite{Witt,Alcoc}. The process
  is accompanied
by a powerful release of energy (of about $10^{53}$ erg) since
nascent light quarks acquire relativistic temperature
\cite{Alcoc,Olinto}.
 Cooling of high-temperature strange stars has been
 proposed as a mechanism of intensive neutrino and
$\gamma$-ray emission (see \cite{Hat,Usov}, and references therein).
There is mounting, albeit controversial, observational evidence for
the existence of strange stars. Likely candidates to be strange
stars are the the compact objects 3C58
 and RX J1856.5-3754.\cite{Helfand,Drake}.
 The possibility to detect indirectly strange quark stars (that  may be
formed after SNe explosions)  by their effect on the SN-ejecta and
connections between  the formation of strange stars and GRBs were
discussed in \cite{Ouyed1,Ouyed2}. Recently, the authors of
paper~\cite{Haen} have pointed out that a soft gamma ray
precursor, like GRB980425/SN1998bw, detected by Swift could
provide evidence for the formation of a quark star.

 Hydrodynamical considerations
lead to the conclusion that the conversion is likely to proceed in
the form of combustion with a well defined front \cite{HB,Ben,TNGF}.
  General conditions under which
combustion in neutron stars may occur have been discussed  in
paper
 ~\cite{Ng}. Based on the conservation laws of hydrodynamics
 the authors of that paper have
investigated the necessary  conditions for  combustion for a wide
ranges of values for bag constant $B$  of the quark matter
equation of state   and the density of the neutron matter (NM).
They have argued that neither slow combustion (deflagration) nor
detonation may be realized. Their analysis, however,
assumed that the energy per baryon must be smaller in the quark
matter after the conversion. This assumption is correct only if
one ignores the kinetic and internal energies gained by the quark
at the combustion front.

In this paper we show that all combustion modes are possible and
present the relevant constraints on the parameters of the neutron
and quark matter equations of state.
\section{Conservation laws}
Our strategy is to assume the existence of a combustion front and
then assess the necessary conditions for its preservation and the
realization of specific combustion modes. Let $v_i$, $n_{i}$, $p_i$,
and, $\varepsilon_i$ be, respectively,
 the  velocity relative to the front, the baryon number density,
  the pressure, and energy density, where the subscript $``i''$ is either
 $``n"$ or $ ``s"$  referring to NM
or strange quark matter (SM), respectively. In a frame of reference
moving with the combustion front, the state of NM and SM at both
sides of the front are related by the equations of baryon flux
conservation,
\begin{equation}
 \label{cond3}
{n_s}{v_s}{\gW_s}={n_n}{v_{n}}{\gW _{n}},
\end{equation}
{\rm the\, momentum\, flow\, conservation,\begin{equation}
\label{cond1} \omega_s v_s  \gW_s ^{2} = \omega_{n} v_{n}  \gW
_{n}^{2};
\end{equation}
 and  energy flow conservation,
\begin{equation}
\label{cond2}
 \omega_s v_s^{2}  \gW_s ^{2} + p_s = \omega_{n}
v_{n}^{2} \gW _{n}^{2}+p_n,
\end{equation}
 where
$\omega_i=p_i + \varepsilon_i $ is the  specific enthalpy and
$W_i=1/\sqrt{1-v_i^2}$ is Lorentz factor~\cite{Land}. These
conservation laws allow us to express the velocities $v_{n}$ and
$v_{s}$ in terms of the energy density and pressure of NM and SM
as follows,
\begin{equation}
v_s^2=\frac{(p_s-p_n)(\varepsilon_n+p_s)}{(\varepsilon_s-\varepsilon_n)(\varepsilon_s+p_n)},\qquad
\label{eq:vels}
\end{equation}
\begin{equation}
v_n^2=\frac{(p_s-p_n)(\varepsilon_s+p_n)}{(\varepsilon_s-\varepsilon_n)(\varepsilon_n+p_s)}.
\label{eq:veln}
\end{equation}
The conservation laws also yield,
\begin{equation}
n_s^2=n_n^2\frac{(\varepsilon_s+p_s)(\varepsilon_s+p_n)}{(\varepsilon_n+p_n)(\varepsilon_n+p_s)}
\, . \label{eq:den}
\end{equation}

\section{Equations of State}
 To make progress, we have to adopt specific forms for the equations of
 state of NM and SM.
We will assume the  ideal gas approximation for the quark SM with
temperature $T$ and chemical potential $\mu$, with vanishing quark
masses $m_q=0$ and a  strong coupling constant $\alpha_s=0$. The
MIT bag model equation of state (EOS)~\cite{cleim} for this matter
is expressed as follows\footnote{\footnotesize For the SM, $T$,
$\mu$ are measured in $\rm fm^{-1}$ and $\varepsilon$ and $P$ in
$\rm fm^{-4}$, while for the NM $\varepsilon $ and $p$ are in $\rm
Mev\; fm^{-3}$. },
\begin{eqnarray}
 \varepsilon_{s} = {\frac{{19}}{{12}}}\pi ^{2}T^{4} +
{\frac{{9}}{{2}}}T^{2}\mu ^{2} + {\frac{{9}}{{4\pi ^{2}}}}\mu ^{4}
+ B,\, \nonumber\\
p_s = \frac{19}{36}\pi ^{2}T^{4} +\frac{3}{2}T^{2}\mu
^{2}+\frac{3}{4\pi ^{2}}\mu
^4 - B;\nonumber \\
n_s = T^{2}\mu + {\frac{{1}}{{\pi
^{2}}}}\mu ^{3},\qquad\qquad\qquad\qquad\nonumber\\
\label{eos1}s_s = {\frac{{19}}{{9}}}\pi ^{2}T^{3} + 3T\mu
^{2},\qquad\qquad\qquad\;
\end{eqnarray}
where $B$ is the MIT bag constant.
 The sound speed in such matter
is defined as $(\partial p/
\partial \varepsilon )_{s} \equiv c_{s}^{2}$.
It is easy to see that the sound speed is $c_{s} = 1 / \sqrt {3}
$.
\begin{figure}[h]
\includegraphics[width=75mm,height=65 mm]{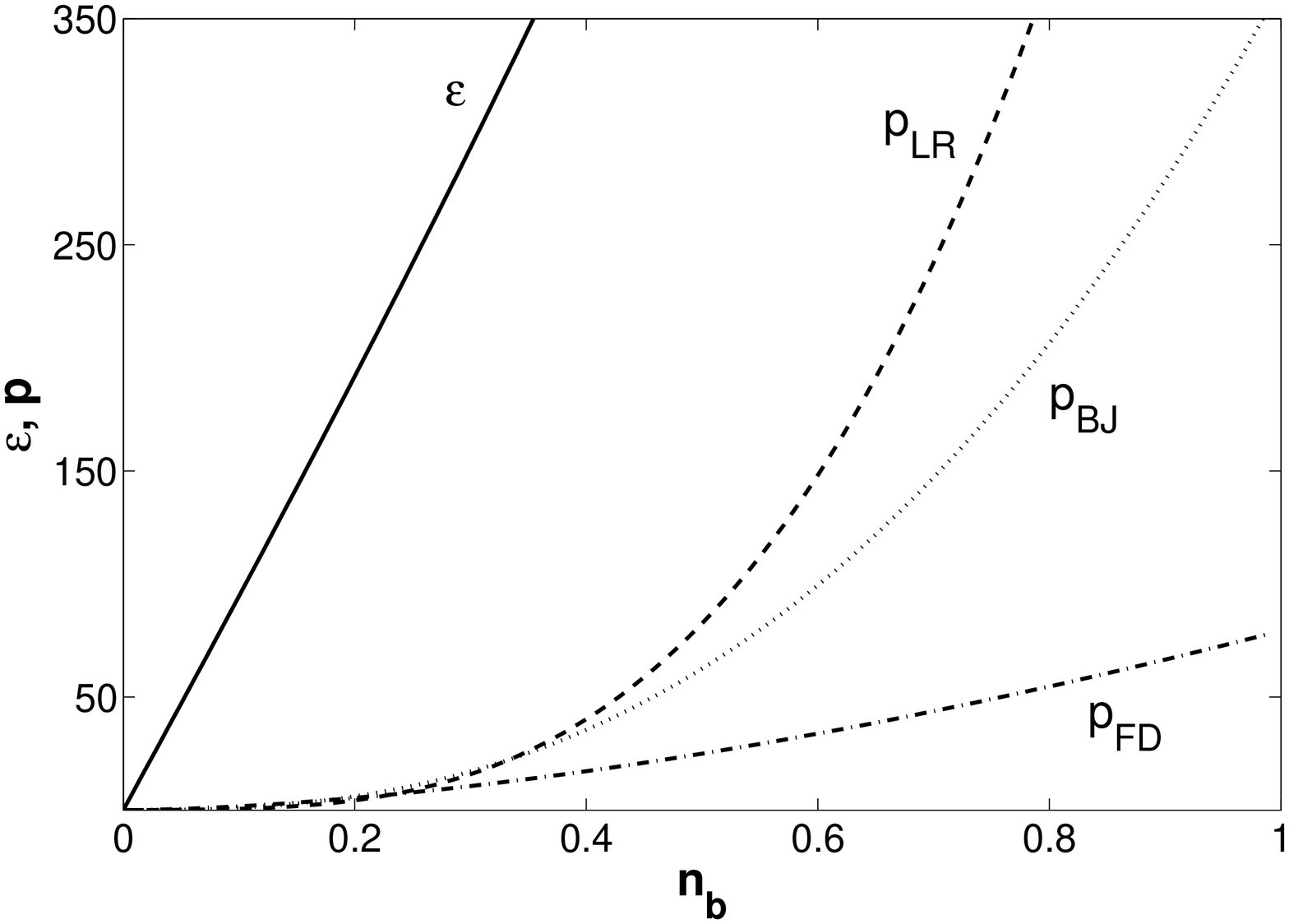}
\caption{{\footnotesize Dependence of the energy density and
 the pressure on the baryon density of zero-temperature NM for
 the  FD  (dott-dashed line),
 BJ  (dotted line), and the LR (dashed line) equations of state.}}
 \label{feos}
 \end{figure}

For the NM we will work with the following three forms  for the EOS,
\begin{itemize}
\item the zero-temperature ideal Fermi-Dirac (FD) neutron gas EOS~\cite{ShapT}
 \begin{equation}
\varepsilon_n =m_n\,n_n+119\,n_n^{5/3}\,,
  \end{equation}
  \begin{equation}
p_n =79.3\;n_n^{5/3}\,;
 \end{equation}
 \item the zero-temperature Bethe-Johnson (BJ) EOS~\cite{ShapT}
  \begin{equation}
  \varepsilon_n=236\;n_n^{2.54}+m_n\,n_n\,,
    \end{equation}
  \begin{equation}p_n=364\;n_n^{2.54}\,;
   \end{equation}
   where $m_n$ is neutron mass in MeV.
   \item finite temperature Lattimer - Ravenhall (LR) EOS~\cite{LR}
   $$\varepsilon_n=d_1\tau+d_2\tau n_n+
\frac{1}{4}d_3 n_n^{2}+d_4 n_n^{3},
  $$
  $$p_n=\frac{2}{3}d_1\tau+\frac{5}{3}d_2\tau n_n+ \frac{1}{4}d_3
n_n^{2}+2d_4 n_n^{3},
   $$
    \begin{equation}
   n_n=\frac{1}{2 \pi^2}\left(\frac{2m^{\star}}{\hbar\beta}
   \right)^{3/2}F_{1/2}(y),
   \end{equation}
 where $\tau$  is the baryonic kinetic energy density
\begin{equation}\tau=\frac{1}{2 \pi^2}\left(\frac{2m^{\star}}{\hbar^2\beta}\right)
   ^{5/2}F_{3/2}(y)\; ,
\end{equation}
   $m^{\star} $ is effective mass of the neutron given by
   \begin{equation}   \hbar^2/2m^{\star}=d_1+d_2n_n\; ,
\end{equation}
   and $F_i(y)$ is Fermi integral,
\begin{equation} F_{i}(y)=\int_0^{\infty}\frac{u^idu}{\exp(u+y)+1}\; .
 \end{equation}
 The coefficients have the values $d_1=20.75\, {\rm Mev\, fm}^2,\;
 d_2 =-8\,{\rm Mev\,fm}^5,\, d_3=-752.27\,{\rm Mev\, fm}^3,$ $d_4 = 466.6\,
 {\rm Mev\, fm}^6;\;  \beta=1/k_BT$.
\end{itemize}
Fig.\ref{feos} is an  illustration of these three choices
for the  NM EOS.

\section{Constraints}

For a given set of the the parameters, $T_{n}$, $T$, $n_{n}$, and
$B$, the  EOS of the NM uniquely determines  $\varepsilon_{n}$ and
$p_{n}$, while the SM EOS  can be used to obtain $\varepsilon_{s}$
and $p_{s}$ as a function of $\mu$. Further, by substitution in the relation
(\ref{eq:den})
 the value of strange matter chemical potential $\mu$ may be  derived.
 Typically,
 there are
several roots for (\ref{eq:den}), but only one  satisfying the
physical condition $\mu>0$. Once $\mu$ is obtained the relations
(\ref{eq:vels}) and (\ref{eq:veln}) yield $v_s^2$ and $v_n^{2}$,
respectively. For any given $T_{n}$, $T_{s}$, $n_{n}$, and $B$
combustion is possible whenever $v_{n}^{2}$ and $ v_{s}^{2}$ are
positive and less than unity. Once this occurs three modes of
combustion are identified as follows:
\begin{figure*}[t]
\includegraphics[width=130mm,height=75 mm]{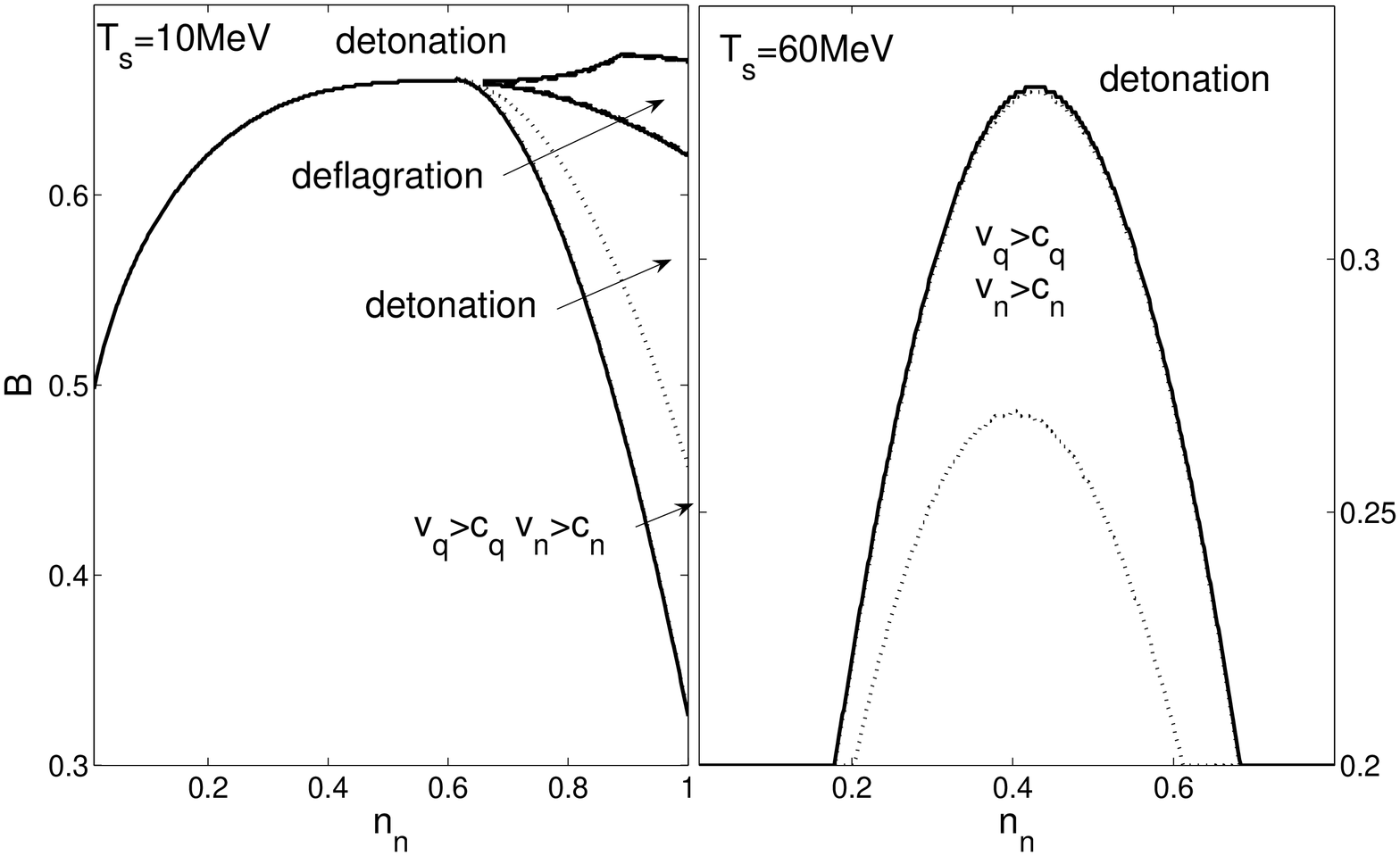}
\caption{ {\footnotesize The different regions of combustion of FD
NM into SM with $T_s=10$ and $T_s=60$ MeV as indicated in the
panels. The regions of the detonation, ``super-sonic'' combustion
(from the viewpoint $v_s>c_s$), and deflagrations are bounded by
 solid lines,  dotted  line, and dashed lines, respectively.
 Regions which are not explicitly
 labeled represent physically unacceptable solutions.}}
 \label{fig1}
\end{figure*}
\begin{itemize}
\item detonation: $p_s>p_n$, $c_n\le v_n \le 1$ and $v_s \le c_s$;
\item deflagration: $v_n<c_n$ and $p_s<p_n$.
\item ``super-sonic'' combustion$p_s>p_n$, $c_n\le v_n \le 1$ and $c_s<v_s<1 $
\end{itemize}
Figs.~(\ref{fig1})-(\ref{fig3}) illustrate the regions in the $n_{n}-B$ plane
in which the  various combustion modes occur, for a few representative
values of $T_{s}$.
 Fig.\ref{fig1} shows results for the FD EOS.
 We can see that all modes of combustion may be realized for
 a wide range of the parameters. Regions which are not explicitly
 labeled represent physically unacceptable solutions having any of the following
 cases: imaginary
 baryon flux; $v_n<c_n,\; v_q>c_n$
  (absolutely unstable combustion); $p_s <p_n$ and $v_n>c_n$; $p_{s}>p_{n}$ and $v_n<c_n$
  (see ~\cite{Land}).

 Fig.~\ref{fig2}
 presents the combustion regions   for BJ
EOS for different temperatures. We can see again conditions for
deflagration, detonation  and ``super-sonic''  combustion. Results
for LR EOS are shown in Fig.\ref{fig3}. They only slightly  differ
from those of BJ EOS.
\begin{figure*}[t]
\includegraphics[width=130mm,height=75 mm]{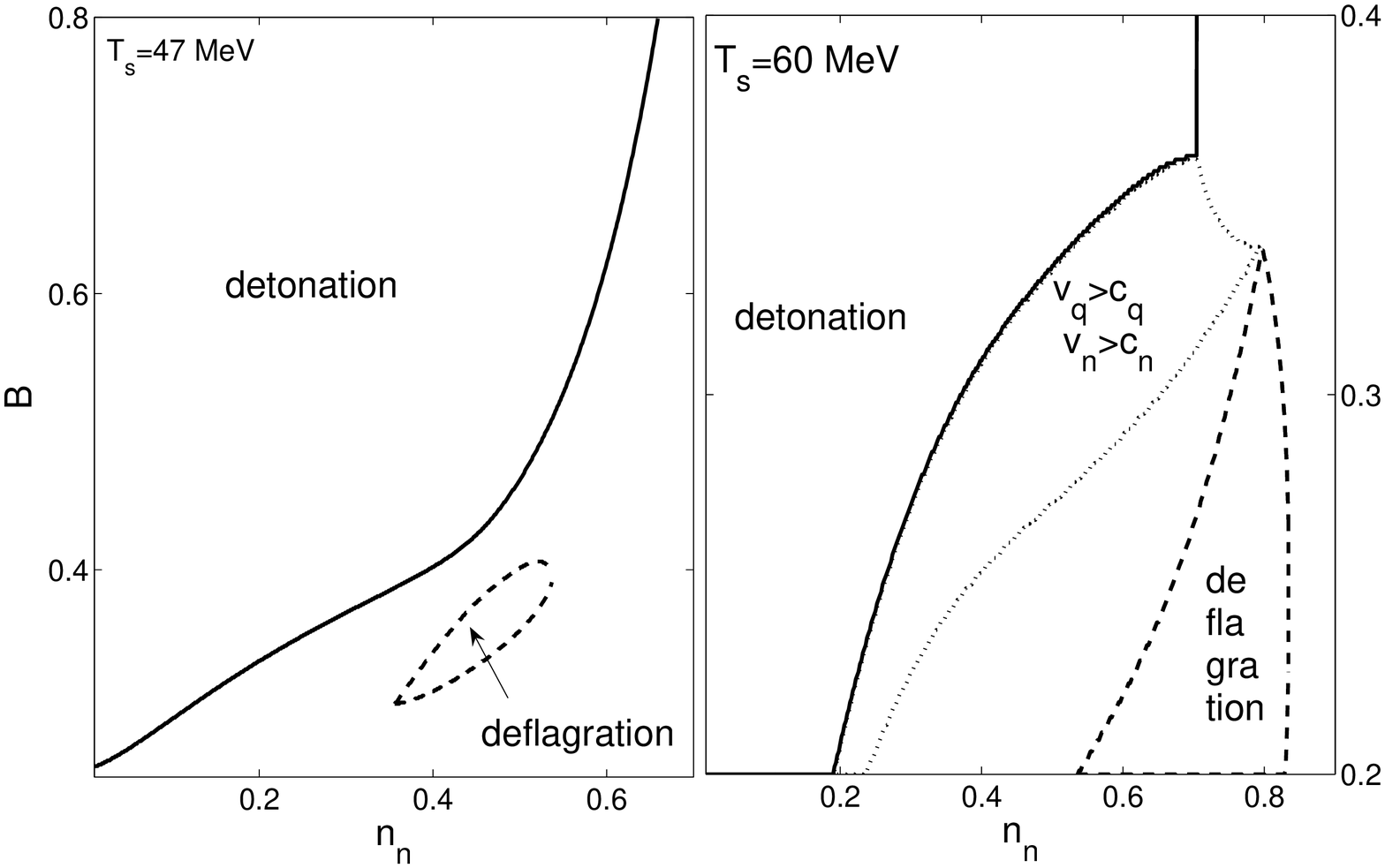}
\caption{ Combustion regions for  BJ EOS NM $T_s=47$ (panel to the left)
and $T_{s} =60$  MeV (to the right). The
notation of the lines is the same as in the previous figure.}
 \label{fig2}
\end{figure*}
\begin{figure*}[t]
\includegraphics[width=130mm,height=75mm]{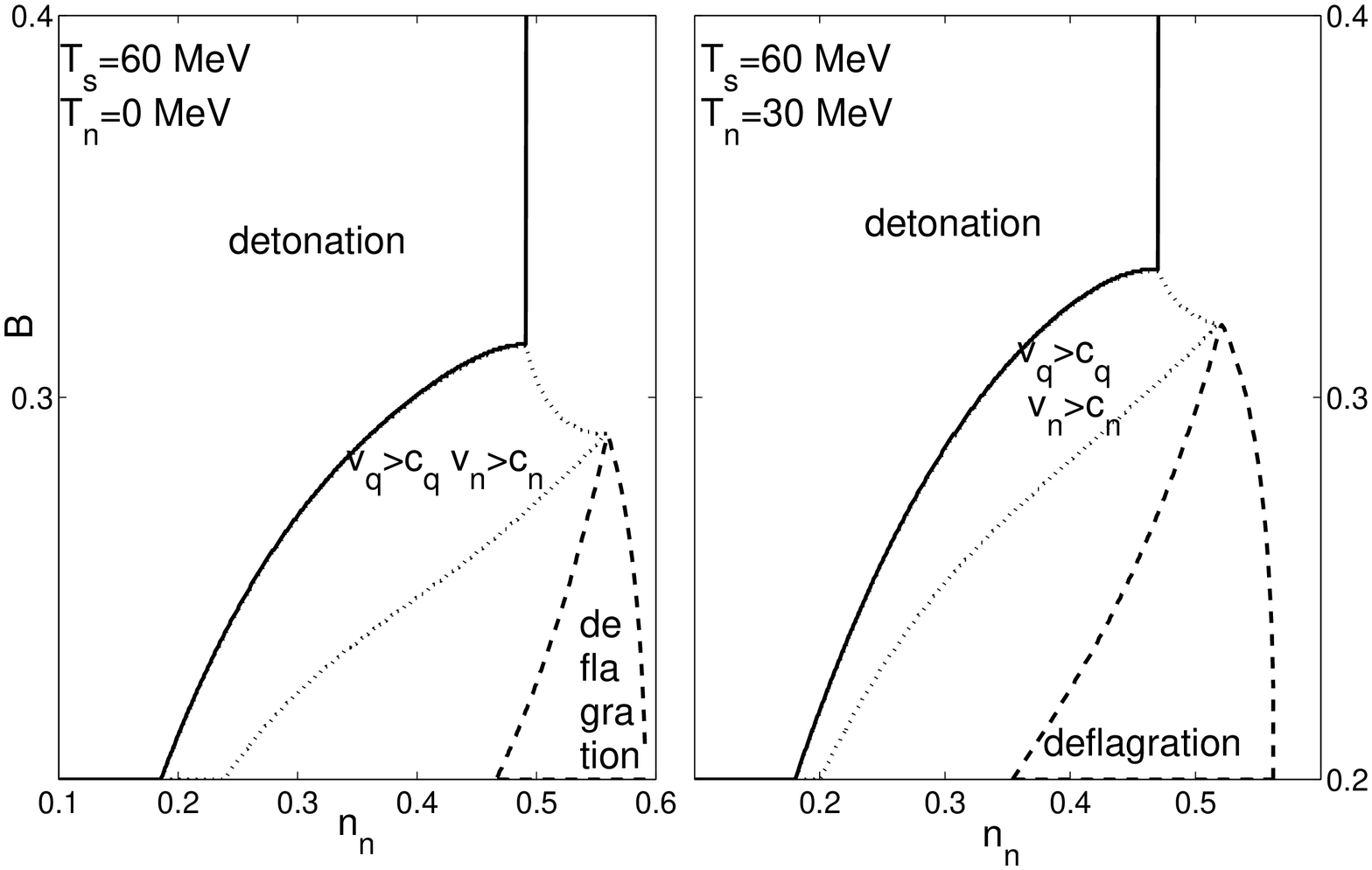}
\caption{{\footnotesize  The combustion regions
for the  LR EOS with  $T_{s}=60$  Mev and $T_{n}=0$ (panel to the left) and
$T_{n}=30$ Mev (to the right).
The notation of the line is the same as in the two previous figures. }}
 \label{fig3}
\end{figure*}

\section{Discussion}
We have shown that
 conversion of neutrons
into strange quark matter  in a combustion mode with a well defined front
 is possible for
a wide range of the parameters, $T_{s}$, $n_{n}$, $T_{n}$, and $B$.
 The
concrete type of the process  depends on the values of these parameters and the
 equation of state of neutron. The results of the current work
differ from   paper~\cite{Ng} which assume that  the energy per
baryon in the strange matter ($\varepsilon_s/n_s$) must be less than
the energy per baryon ($\varepsilon_n/n_n$) in neutron matter.
However,  the energies $\varepsilon_{s}$ and $\varepsilon_{n}$
contain thermal contributions in addition to the rest energies of
the baryons and therefore the condition,  $\varepsilon_n/n_n>
\varepsilon_s/n_s$, does not necessarily  hold in general.

In the current paper we have treated $T_{s}$ as a free parameter.
Physically, the value of $T_{s}$ is affected by the details of the
initial conditions for the conversion process. In a subsequent
paper, we will discuss in detail, possible constraints on this
parameter assuming that seed strange matter is initially generated
by means of quantum nucleation~\cite{Bomb1,Bomb}.
\section{Acknowledgements}
I.T. thanks V.Gurovich for useful
discussions.

\end{document}